\documentclass[prl,twocolumn,showpacs,preprintnumbers,amsmath]{revtex4}

\usepackage{graphicx}
\usepackage{dcolumn}
\usepackage{bm}
\usepackage{color}

\begin{document}
\newcommand{\js}[1]
{
\textcolor{red}{{#1}}
}

\preprint{APS/123-QED}

\title{Knudsen gas provides nanobubble stability}
\author{James R. T. Seddon$^1$}
\email{j.r.t.seddon@utwente.nl}
\author{Harold J. W. Zandvliet$^{2}$}
\author{Detlef Lohse$^1$}
\affiliation{$^1$Physics of Fluids,  $^2$Physics of Interfaces and Nanomaterials,  MESA+ Institute for Nanotechnology, University of Twente, P.O. Box 217, 7500 AE Enschede, The Netherlands}
\begin{abstract}
We provide a model for the remarkable stability of surface nanobubbles to bulk dissolution. The key to the solution is that the gas in a nanobubble is of Knudsen type. This leads to the generation of a bulk liquid flow which effectively forces the diffusive gas to remain local. Our model predicts the presence of a vertical water jet immediately above a nanobubble, with an estimated speed of $\sim3.3\,\mathrm{m/s}$, in good agreement with our experimental atomic force microscopy measurement of $\sim2.7\,\mathrm{m/s}$. In addition, our model also predicts an upper bound for the size of nanobubbles, which is consistent with the available experimental data.
\end{abstract}
\pacs{}
\maketitle

Classical diffusion predicts a lifetime of $\sim 1\,\mathrm{\mu s}$ for a nanoscopically-sized bubble.  So the fact that surface nanobubbles~\cite{parker1994,lou2000,tyrrell2001,holmberg2003,steitz2003,borkent2007,yang2008,ralston2010,craig2010,seddon2011} (typical height $\sim 20\,\mathrm{nm}$ and width $\sim 100\,\mathrm{nm}$) persist for \textit{at least} $11$ orders of magnitude longer than this~\cite{zhang2008} is both remarkable and puzzling.  Are classical diffusion laws simply not applicable at these length scales? Are nanobubbles coated with diffusion-limiting molecules~\cite{ducker2009,das2010}? Or does the gas indeed diffuse out, but is balanced by an equivalent influx~\cite{brenner2008}?   Supersaturation was thought to be the key to nanobubble nucleation and stability~\cite{zhang2005,yang2007}, but it is now known that this is not a requirement~\cite{seddon2010c}.  It was also originally thought that nanobubbles may not actually contain gas~\cite{evans2004}, but this is not correct either~\cite{zhang2007}.  This is one of the outstanding questions in fluid dynamics~\cite{ball2003,craig2010,seddon2011}.

Surface nanobubbles are fundamentally interesting.  For example the gas-side contact angle that they make with the solid is always very low, regardless of the substrate chemistry, and  also depends on size~\cite{simonsen2004,zhang2006,borkent2010,seddon2011a}.  This is  in clear contrast to the classical view that the  contact angle is a material property and should be substrate dependent and size independent.  Nanobubbles also have clear potential in applications, such as controlling slip in microfluidic devices~\cite{wang2009b,wang2010} and surface cleaning in nanofabrication processes~\cite{switkes2003,wu2008}.  Hence, understanding their stability is paramount.

In this Letter we suggest a solution to the mystery of nanobubble stability by demonstrating that the exact nature of the gas, i.e. Knudsen, is the key.  The symmetry-broken geometry provided by the hard substrate and the `leaky' liquid/gas interface thus generates a bulk gas flow.  In turn, due to the continuity of shear stress boundary condition, this bulk gas flow leads to a bulk liquid flow and, due to conservation of mass, the gas-rich liquid circulates from the bubble apex around to the three-phase line. Hence, although the gas molecules do indeed dissolve into the liquid as  expected, they remain local to the bubble in the gas-rich liquid stream and are effectively transported back to the three-phase line for re-entry.  We  validate our theory by performing  non-contact-mode  open-loop atomic force microscopy in the liquid environment, which has enabled us to measure an incredible $2.7\,\mathrm{m/s}$ upthrusting water jet immediately above a nanobubble, in good agreement with the $3.3\,\mathrm{m/s}$ jet predicted by our model.  Hence, our measurements clearly demonstrate that surface nanobubbles are in a dynamic equilibrium.

Knudsen gases differ from their classical counterparts insomuch as the molecules hardly interact with each other.  Thus, rather than a test volume possessing zero preferred direction, as would be the case for an ideal gas, Knudsen gas molecules mainly travel due to energy exchange with the walls, so are heavily dependent on the geometry of their surroundings.    Thus, the symmetry-broken geometry offered by the hard substrate and liquid/gas interface of a Knudsen gas-filled surface bubble leads to more gas traveling away from the solid substrate than is reflected back from the liquid/gas interface (i.e. some gas molecules are transmitted and diffuse away).  Hence, the gas flow mimics the broken symmetry of the geometry such that every volume element of gas  possesses a bulk flow away from the substrate.  This is a generic statement for any Knudsen gas with thermal drive and one leaky wall (for nanobubbles the thermal drive comes from the substrate, which we treat here as a heat bath).

The requirement for Knudsen gas behaviour is dependent on the Knudsen number, $Kn$, i.e. the ratio between the molecular mean free path, $\lambda$, and the typical length scale of the container.  For a surface bubble, this length scale is the bubble height, $h$, so Knudsen gas behaviour is exhibited when
\begin{equation}\label{eq:kn}
Kn = \frac{\lambda}{h} = \frac{kT}{\sqrt{2}\sigma R(p_0+2\gamma/R)}\frac{1}{(1-\cos\theta)}>1,
\end{equation}
where $kT$ is the thermal energy,  $R$ the radius of curvature, $\gamma$ the surface tension, $\sigma$ the molecular collisional cross-section, $p_0$ the ambient pressure, and $\theta$ the gas-side contact angle.  For macroscopic bubbles $R\gg 2\gamma / p_0$ so $Kn\ll 1$ and the gas is ideal, but for nanoscopic bubbles $R\ll 2\gamma / p_0$  and the requirement for Knudsen gas behaviour becomes surprisingly only dependent on the contact angle.  As an example, an oxygen nanobubble on graphite at room temperature will exhibit Knudsen gas behaviour if the contact angle is $\theta \lesssim 25\,\mathrm{^o}$.

Coincidentally, an as yet unexplained mystery of surface nanobubbles is that their contact angles are not consistent with their micro- or macroscopic counterparts.  Instead, their contact angles are always found in the range $5\,\mathrm{^o}\lesssim \theta \lesssim 25\,\mathrm{^o}$ for hydrophobic surfaces, regardless of the substrate chemistry~\cite{simonsen2004,zhang2006,borkent2010,seddon2011a}, suggesting that their internal gas is \textit{always} of Knudsen type.  In what follows we shall demonstrate that this puzzling observation may be the key to the nanobubbles' mysterious stability.

\begin{figure}
\begin{center}
\includegraphics[angle=0,width=7.5cm]{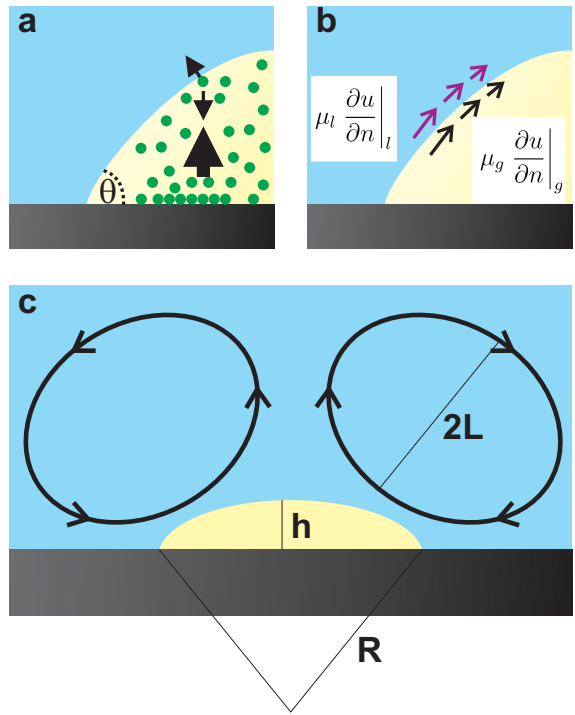}
\end{center}
\caption{\label{fig:cartoon}Knudsen gas streaming and nanobubble geometry.  (a) The broken symmetry created between one solid surface and one `leaky' liquid/gas interface leads to bulk upwards flow in the Knudsen gas. (b) The tangential component of the bulk Knudsen gas flow drives a bulk liquid flow due to the continuity of shear stress at the liquid/gas interface. (c) Finally, due to conservation of mass, this gas-rich liquid stream circulates upwards at the bubble apex and back around to the three-phase line, effectively transporting the diffusive outfluxing gas back to the three-phase line for re-entry \cite{brenner2008}.}
\end{figure}

We begin with the gas molecules arriving at the liquid/gas interface.  These molecules have not been able to interact with each other, due to their Knudsen behaviour, so they still possess the directional flow away from the substrate (Figure \ref{fig:cartoon}a).  At the apex of the nanobubble they either reflect back into the bubble or transmit and dissolve.  However, everywhere else on the liquid/gas interface the gas molecules arrive, on average, non-normal, so their bulk velocity can be decomposed into both a normal and a tangential component.  The normal component acts like at the apex (reflection or transmission), but it is the \textit{tangential} component that we are interested in.  If a tangential velocity component exists in the gas phase it always points towards the apex of the bubble.  This then communicates with the liquid phase through the assumed continuity of shear-stress boundary condition (Figure \ref{fig:cartoon}b), i.e.
\begin{equation}
\mu_g \left. \frac{\partial u}{\partial n} \right|_g = \mu_l \left. \frac{\partial u}{\partial n} \right|_l,
\end{equation}
where $\mu$ is the viscosity and $\frac{\partial u}{\partial n}$ is the tangential velocity gradient of the gas, $g$, or liquid, $l$, phase.  In this picture we have treated the system as a continuum (we are using $Kn=1$, which means the gas is $50\,\%$ ideal and $50\,\%$ molecular, both interpretations are correct) -- this macroscopic assumption is at least partially valid but it must eventually breakdown.  (The microscopic description is complimentary: The gas molecules in the bubble have a net upflux, but they must travel radially outwards once they dissolve in the liquid.  In order to change direction at the liquid/gas interface they transfer momentum to the liquid molecules in the direction from the three-phase line towards the apex.)  Hence, the upward flow of the Knudsen gas in the nanobubble induces a bulk flow in the adjacent liquid.  The strength of the drive is dependent on the precise position on the liquid/gas interface, with maximum effect near the three-phase line and zero effect at the apex.  The importance of this result is shown in Figure \ref{fig:cartoon}c: The flow in the liquid  tangential to the liquid/gas interface creates a local circulatory stream due to mass conservation.  (Note that \textit{diffusive} gas flow can also lead to liquid streaming in the extreme limit of inviscid flow~\cite{acrivos1962}.)

\renewcommand{\baselinestretch}{1}

The appropriate scaling of the shear stress is the molecular speed distributed over the bubble length scale, i.e. $\left. \frac{\partial u}{\partial n}\right|_g \sim u_g/R$, whilst in the liquid the induced velocity is distributed over the radius, $L$, of the circulation loop, i.e. $\left. \frac{\partial u}{\partial n}\right|_l \sim u_l/L$.  So, although the gas in a nanobubble does indeed dissolve into the liquid, this gas-rich liquid is streamed at speed $u_l \sim \frac{\mu_g u_g L}{\mu_l R}$ from the bubble apex around to the three-phase line, where it can re-enter through either the attractive potential of the hydrophobic wall~\cite{dammer2006} or through adsorption to the substrate and surface diffusion.

To test our hypothesis of gas-driven streaming in the bulk liquid we have performed non-contact-mode open-loop feedback-disabled  atomic force microscopy in the liquid environment.  This has allowed us to make a direct measurement of the force-field in the liquid, with the AFM cantilever acting as a local force, and thus velocity, probe.  The liquid was ultrapure water (Simplicity 185 system, Millipore, France), which had been thoroughly degassed before supersaturation with $3\,\mathrm{atm}$ of argon gas.  This was then deposited onto freshly cleaved HOPG before scanning in both tapping mode and the non-contact open-loop feedback-disabled mode.  The cantilevers were Au-back-coated Si$_3$N$_4$ Veeco NPG probes (radius of curvature $30\,\mathrm{nm}$, full tip cone angle $35\,\mathrm{^o}$), with resonance frequencies in liquid of $\omega_0^{liq} \approx 15-25\,\mathrm{kHz}$.

\begin{figure}
\begin{center}
\includegraphics[angle=0,width=8cm]{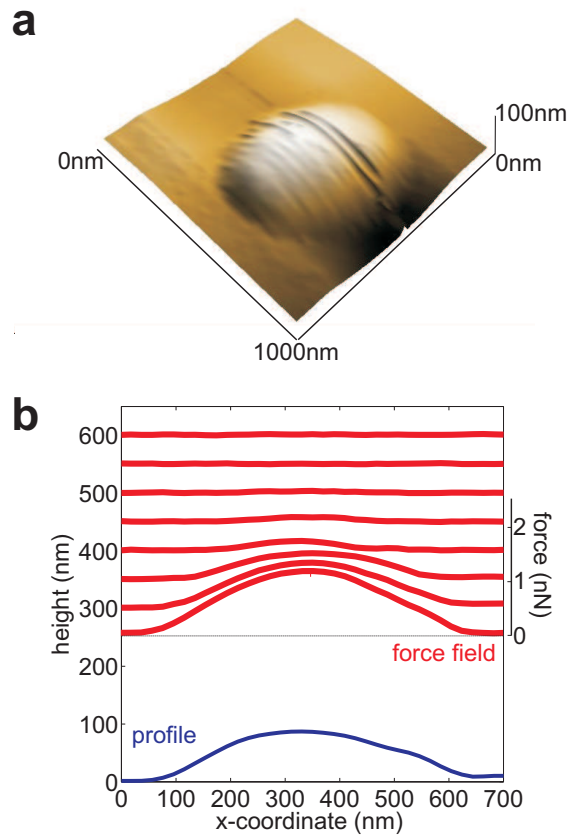}
\end{center}
\caption{\label{fig:force}(a)  Three-dimensional topological image ($1\,\mathrm{\mu m} \times 1\,\mathrm{\mu m} \times 100\,\mathrm{nm}$) of one of the nanobubbles investigated in this study.  The nanobubble was created by depositing room temperature ultra pure (Millipore) water on highly-oriented-pyrolitic-graphite (HOPG), also at room temperature.  The water had been thoroughly degassed before supersaturating it with $300\,\%$ argon, resulting in a  bubble with radius of curvature $R\approx1.4\,\mathrm{\mu m}$ and contact angle $\theta \approx 20\,\mathrm{^o}$.   The nanobubble was topologically imaged regularly over $\sim 12\,\mathrm{hrs}$ to confirm that its size was not changing with time. (b) Cross-sectional line scan (blue) of the nanobubble in (a), with  feedback-disabled non-contact-mode force field measurements (red), taken at $250\,\mathrm{nm}$ to $600\,\mathrm{nm}$ above the substrate, in steps of $50\,\mathrm{nm}$.  At $250\,\mathrm{nm}$ from the substrate ($\sim160\,\mathrm{nm}$ above the nanobubble's apex) the AFM cantilever was deflected upwards by a $\sim 1\,\mathrm{nN}$ force, which decreased in magnitude with increasing cantilever-substrate separation until it became smaller than the AFM resolution ($\sim5\,\mathrm{pN}$) at $550\,\mathrm{nm}$.  The force ordinate on the right-hand side of the image is for the bottom-most force field measurement.  The proceeding force field measurements have the same scale, but with their zeroes offset according to the height at which they were recorded (left-hand ordinate). We recalibrated the AFM cantilever to the substrate between each force field measurement in order to maintain accurate separation with minimal drift for each scan.}
\end{figure}

The results of our measurements are shown in Figure \ref{fig:force}b, where the blue curve is a line scan over the apex of the nanobubble in Figure \ref{fig:force}a, and the red curves are measures of the force field taken at various distances above the substrate (from $\sim250\,\mathrm{nm}$ to $\sim600\,\mathrm{nm}$, in steps of $\sim50\,\mathrm{nm}$).  A clear upthrust exists in the immediate vicinity of the nanobubble.  The maximum force exerted by this flow was found to be $\sim 1\,\mathrm{nN}$ at a distance of $250\,\mathrm{nm}$ above the nanobubble, with the flow still measurable as far away as $\sim 500\,\mathrm{nm}$.

(To test whether the upthrust force was electrostatic, we changed the bias of the cantilever with respect to the sample from $-1.5\,\mathrm{V}$ to $+1.5\,\mathrm{V}$ in steps of $0.5\,\mathrm{V}$.  However, no change in force was evident.  We also investigated the effect of different ionic concentrations within the liquid by using $0.00\,\mathrm{M}$, $0.01\,\mathrm{M}$, $0.02\,\mathrm{M}$, $0.05\,\mathrm{M}$, $0.10\,\mathrm{M}$, $0.20\,\mathrm{M}$, and $0.50\,\mathrm{M}$ NaCl solutions.  Again, no change in the upthrust force was detectable.  Thus, the upthrust forces measured from nanobubbles are not electrostatic.)

The implication is clear: Our experimental observations demonstrate that nanobubbles are not in a true equilibrium.  This settles the ongoing  debate in the field~\cite{brenner2008,ducker2009}: The system is in a \textit{dynamic equilibrium}.

We benefitted greatly in our experiments from the fact that the upthrusting jet was \textit{focussed} immediately above the nanobubble.  Due to the continuity of mass, a central up-flow must be balanced with a downwards flow in a circulatory stream.  However, we were  not able to measure this downwards motion of the circulation stream, presumably because it is averaged out over a very large annular area. (The prediction from the continuity of flux is that the downwards velocity is $\sim1\,\%$ that of the upwards jet, where we have taken a fixed radius circulatory stream of $50\,\mathrm{nm}$ rising upwards immediately above the nanobubble and returning back downwards at a radial position of $2L$.)

We can validate our model further by measuring the velocity of the jetting water.  Treating the nose of the AFM cantilever as a sphere of radius $r$, the effective fluid velocity can be estimated by equating our $\sim 1\,\mathrm{nN}$ measurement with  Stokes' drag, i.e. $u_l = F/6\pi \mu r$.  In the case of the nanobubble in Figure \ref{fig:force}, the water jet stream induced by the nanobubble's Knudsen gas travelled at an incredible $2.7\,\mathrm{m/s}$!  This is in good agreement with the $u_l \sim \frac{\mu_g u_g L}{\mu_l R}= 3.3\,\mathrm{m/s}$ prediction of our model, where we have used $L = 600\,\mathrm{nm} - 90\,\mathrm{nm}$ as the radius of the circulation stream (the limit of the circulation loop measured in Figure \ref{fig:force} was $600\,\mathrm{nm}$ above the substrate, and we have subtracted the $90\,\mathrm{nm}$ height of the  nanobubble -- the precise limit is expected to be between $L$ and $2L$).

We must clearly consider heat generation from this exceptional jet.  Heat generation is through viscous dissipation within the liquid, and has rate $\epsilon_{diss} = \mu_l \left. \frac{\partial u}{\partial n} \right|_l^2$.  Hence, the rate of heating is
\begin{equation}
\Delta T/\Delta t = \frac{\mu_l}{c\rho} \left. \frac{\partial u}{\partial n} \right|_l^2,
\end{equation}
which is an incredible $10^4\,\mathrm{K/s}$ (where $\rho$ and $c$ are the water density and specific heat capacity, respectively).  However, we clearly do not see such a large increase in temperature in the experiments.  There are two possible explanations: (i) Firstly, the heat is generated in the small volume of liquid that forms the recirculation loop, but this is advected away through the entire droplet.  This reduces the temperature increase by the ratio of these two volumes, i.e. approximately $10^{-12}\,\mathrm{K}$ in our experiment. (ii) Secondly, the relevant time scale for the rate of heating is the travel time of one loop of the circulatory stream, $t = 2\pi L/u_l$.  Given that we are treating our solid surface as a heat bath, and that the solid surface is much more efficient at conducting the heat away than the liquid, once the heated liquid has circulated around to the wall it can efficiently exchange energy and return to temperature $T$.  The estimated temperature increase of the liquid using this time scale is $\approx10\,\mathrm{mK}$.  In either case, dissipative heating is negligible.

Finally, our model predicts an upper limit for the size of a nanobubble.  We have chosen to use $Kn = 1$ as the limiting factor for the stabilising effect in this Letter.    If the height of the nanobubble was larger than the mean free path of gas in \textit{atmospheric} conditions ($\lambda_0\approx 100\,\mathrm{nm}$), we may expect the Knudsen gas behaviour to break down.  All nanobubble studies to date have had nanobubbles with heights smaller than this.

To summarise: (i) Surface nanobubbles contain Knudsen gas which possesses a bulk flow due to the thermal energy of the substrate and the leaky liquid/gas interface; (ii) this bulk gas flow drives a bulk liquid flow due to the continuity of shear stress; (iii) the gas-rich liquid flow is a circulatory stream from the apex to the three-phase line, due to the conservation of mass; (iv) the gas re-enters the nanobubble at the three-phase line, replenishing the diffusive outflux.

We acknowledge funding from the European Community's Seventh Framework Programme (\textit{FP7/2007-2013}) under \textit{grant agreement} number 235873 and from the Foundation for Fundamental Research on Matter (FOM), which is sponsored by the Netherlands Organization for Scientific Research (NWO).

\end{document}